\documentclass{PHYEAUTH}
\usepackage{graphicx}
\usepackage{amsmath}
\usepackage{amssymb}

\begin{document}

\begin{frontmatter}

\title{
Cyclotron Resonance Study of the Two-Dimensional Electron 
Layers\\ and Double-Layers in Tilted Magnetic Fields
}

\author{Nataliya A.~Goncharuk},
\author{Ludv\'{\i}k~Smr\v{c}ka \thanksref{thank1}},
and
\author{Jan~Ku\v{c}era} 

\address{Institute of Physics ASCR, Cukrovarnick\'a 10, 162 53 Praha
6, Czech Republic}

\thanks[thank1]{
Corresponding author. 
E-mail: smrcka@fzu.cz}

\begin{abstract}
The far-infrared absorption in two-dimensional electron layers subject
to magnetic field of general orientation was studied theoretically.
The Kubo formula is employed to derive diagonal components of the
magneto-conductivity tensor of two\discretionary{-}{-}{-}dimensional
electron single-layers and double-layers.  The parabolic quantum well
is used to model a simple single-layer system. Both single-layer and
double-layer systems can be realized in a pair of tunnel-coupled,
strictly two-dimensional quantum wells. Obtained results are compared
to experimental data.
\end{abstract}

\begin{keyword}
single-layer \sep double-layer \sep two-dimensional electron system 
\sep cyclotron resonance
\PACS 73.21.Fg \sep 73.63.Hs 
\end{keyword}
\end{frontmatter}

\section{Introduction}
The cyclotron resonance in the absorption of far-infrared radiation in
two-dimen\-sional electron single-layers and double-layers occurs when
its frequency $\omega$ coincides with the frequency $\omega_c$ of the
circular motion of electrons subjected to a magnetic field $\vec
B$. 

In the quasi-classical picture of the cyclotron resonance, the current
(and, consequently, the absorbed power) induced in the layer by the
electric field reach maximum at resonance conditions, $\omega =
\omega_c$.

From the point of view of quantum-mechanics, the cyclotron resonance
is described as excitation of electrons from occupied to unoccupied
Landau levels by photons with the energy $\hbar\omega$ equal to the
energy difference between levels, $\hbar\omega_c = \hbar |e| B/m^*_c$.

The cyclotron mass is defined by $m_c^*=|e|B/\omega_c$.  In strictly
two-dimensional systems only $B_{\perp}$, the field component
perpendicular to the electron layers, contributes to the cyclotron
motion of electrons and a single cyclotron mass characterizes the
cyclotron motion for all magnitudes of the magnetic field $B_{\perp}$.
Both quasi\discretionary{-}{-}{-}classical and quantum-mechanical
approaches should yield the same results for small $B_{\perp}$, when
many Landau levels are occupied.
  
The situation becomes more complicated if we consider
quasi-two-di\-men\-sional electron layers and double-layers in magnetic
fields of a general orientation. Due to the finite width of quantum
wells, the electronic structure is strongly affected by the in-plane
component of the field, $B_{\parallel}$.  The cyclotron mass is
not longer a constant and becomes a function of both $B_{\parallel}$
and $B_{\perp}$, $m^*_c=m^*_c(B_{\parallel},B_{\perp})$, as
demonstrated by recent experiments (see e.g.~[1]).

We have studied this problem theoretically. Assuming the Landau gauge
of the vector potential, $\vec{A}=(-B_{\perp}y +B_{\|}z, 0, 0)$, and a
wave function in the form $\psi \propto \exp{ik_x x}\, \varphi(y,z)$,
the Hamiltonian can be written as
\begin{equation}
\label{hamilton}
H=\frac{1}{2m^*}(\hbar k_x +eB_{\perp}y-eB_{\|}z)^2+\frac{p_y^2}{2m^*}
+\frac{p_z^2}{2m^*}+V_{\rm conf}(z).
\end{equation}

First, parabolic quantum wells are employed to model single-layer
systems.  Then, a pair of tunnel\discretionary{-}{-}{-}coupled,
strictly two-di\-men\-sional quantum wells located at the distance
$d$, is studied.  These simple models allow to carry out most
calculations analytically and yet are able to capture the essence of
physics of single-layer/double-layer systems.

In quantum-mechanical approach, the linear response theory (the Kubo
formula) is used to evaluate diagonal components of the
magnetoconductivity tensor responsible for determination of the
dissipated power. Since our aim is to study mainly the
field-dependence of the energy levels $E_{m,n}=E_{m,n}(B_{\|},
B_{\perp})$ and of the transitions between them, we completely neglect
the broadening of levels by scattering of electrons.

In both single-layers and double-layers we consider only the ground
and excited subbands which yield two fans of Landau levels. Each
Landau level is denoted by a subband index $m$, and a level index
$n$. Corresponding wave functions determine probabilities of
transitions between levels via the dipole matrix elements in the Kubo
formula.
%
%
\section{Results and discussion}
\subsection{Parabolic quantum wells}
The quantum well with a parabolic confining potential
$V_{\rm conf}(z)=m^*\Omega^2z^2/2$ represents a particularly simple
model which can be solved analytically.  The frequency $\Omega$
defines the separation of two subbands at ${\bf B}=0$. Eigenenergies
are given as solutions of two coupled harmonic oscillators with
frequencies depending strongly on $B_{\parallel}$ and $B_{\perp}$. The
quasi-classical results are obtained from quantum-mechanical ones
simply by taking limit $B_{\perp}\rightarrow 0$.

In tilted magnetic fields Landau-level mixing occurs and, besides 
intrasubband transitions, $\delta m=0$, $\delta n=\pm 1$, also
intersubband transitions, $\delta n=0$, $\delta m=\pm 1$, become
bright. For fixed $B_{\perp}$, the separation of mixed  levels
decreases with increasing $B_{\parallel}$.  As it is the same for all
adjacent levels, the cyclotron mass does not depend on the level index
and grows with $B_{\parallel}$ in our simple model.
\begin{figure}[t]
\begin{center}\leavevmode
\includegraphics[width=0.95\linewidth]{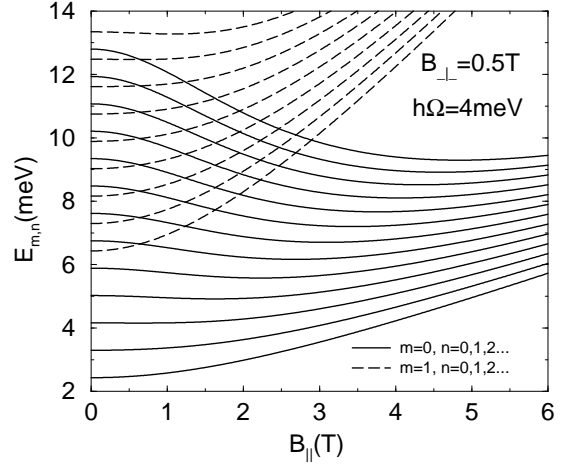}
\caption{Landau eigenenergies as  functions of $B_{\parallel}$ for
fixed $B_{\perp}=0.5~\rm~T$. Results of calculation for the parabolic
model of 2D electron single-layers. The energy of subband separation
is $\hbar\Omega=4~\rm~meV$.}
\label{fig01}
\end{center}
\end{figure}
\begin{figure}[h]
\begin{center}\leavevmode
\includegraphics[width=0.95\linewidth]{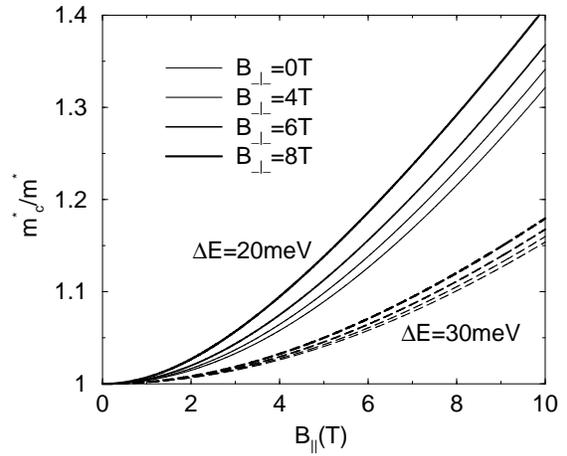}
\caption{The $B_{\|}$-dependence of $m^{\ast}_{c}$ for different
fixed $B_{\perp}$ and $\Omega$. The parameter $\Omega$ determines the
subband separation, $\Delta E = E_{1,n}-E_{0,n} = \hbar\Omega$.}
\label{fig02}
\end{center}
\end{figure}

\begin{figure}[h]
\begin{center}\leavevmode
\includegraphics[width=0.95\linewidth]{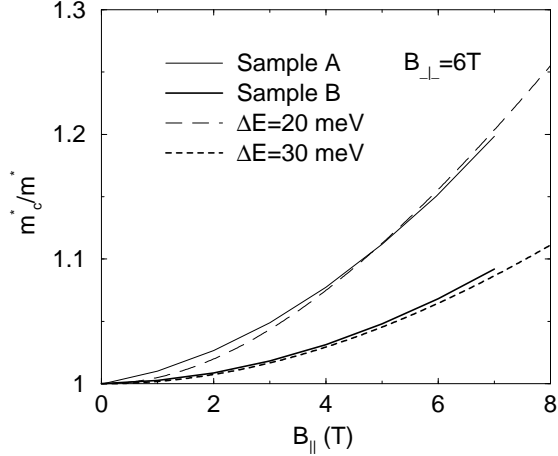}
\caption{The $B_{\|}$ dependence of $m^{\ast}_{c}$ at
$B_{\perp}=6$~T. Theoretical curves (dashed lines) are shown in
comparison with experimental results of S.~Takaoka~\cite{takaoka}
(full lines) on samples with different electron concentrations
($2.0\times 10^{11}~\rm~cm^{-2}$ in sample A and $3.0\times
10^{11}~\rm~cm^{-2}$ in sample B).}
\label{fig03} 
\end{center} 
\end{figure}

Variation of the electron concentration can be modeled by changing
$\Omega$: the higher the concentration, the narrower the well. We have
found that the wider parabolic wells are more sensitive to the
magnetic field as demonstrated in Fig. \ref{fig02}. We have also shown
(see Fig. \ref{fig03}) that with our model one can reasonably fit the
$B_{\|}$- and $B_{\perp}$\discretionary{-}{-}{-}dependence of
experimental data~\cite{takaoka}.
\subsection{Double-well structures}
In double wells, the lowest bound states of individual quantum wells form
symmetric and antisymmetric pairs. The ground and excited states
become separated on energy scale by $2|t|$, where $t$ is the coupling constant
(see e.g. \cite{hu}). 
\begin{figure}[h]
\begin{center}\leavevmode
\includegraphics[width=0.95\linewidth]{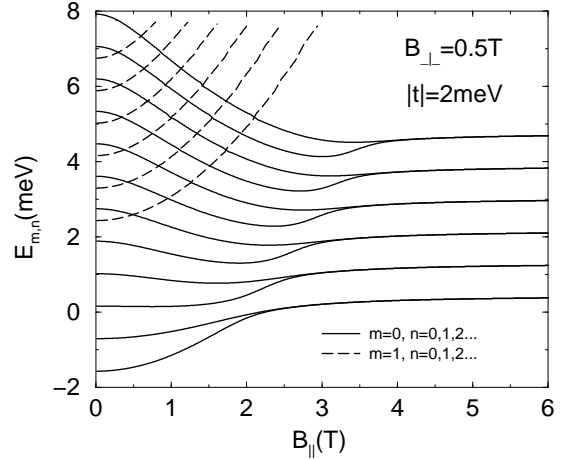}
\caption{Eigenenergies of the 2D electron systems as a function of
$B_{\parallel}$ at fixed $B_{\perp}=0.5$~T and $|t|=2$~meV. Lower and
upper bunch of lines denote Landau levels from the ground (bonding)
and excited (antibonding) subband, respectively.}
\label{fig04} 
\end{center}
\end{figure}
In real space, the bonding and antibonding 2D electron layers overlap
and share the full width of a double-well structure.

The Landau levels with high quantum numbers can be described
quasi-classically~\cite{smrcka,s&t} if $B_{\perp} \ll B_{\parallel}$.
The quantum-mechanical approach must be employed for the case of
strong tilted field: we must solve coupled differential equations for
the left and right components of the double-layer wave function.  The
resulting eigenenergies and eigenvectors are used to calculate the
cyclotron effective mass and transition matrix elements.
\begin{figure}[b]
\begin{center}\leavevmode
\includegraphics[width=0.95\linewidth]{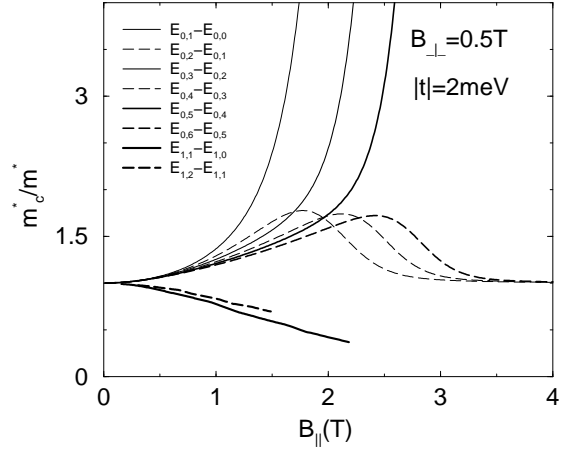}
\caption{Cyclotron effective masses $m^{\ast}_{c}$ 
calculated from energy difference between adjacent Landau levels of 
the band structure spectrum (figure~\ref{fig04}) as a function 
of the in-plane magnetic field.}
\label{fig05}
\end{center}
\end{figure}

The eigenenergies in the excited (antibonding) subband, shown in
Fig. \ref{fig04}, sharply grow with increasing $B_{\parallel}$ for
given $B_{\perp}$. The cyclotron mass corresponding to transition
between Landau levels of the antibonding subband decreases with
increasing in-plane field and differs for each pair of levels 
(see Fig. \ref{fig05}).

The behaviour of levels from the ground subband is more complicated.
The in-plane component of the magnetic field reduces the tunneling
between wells and, at the high magnetic field limit, the single-layer
of bonding electrons is divided into two layers, each layer being
localized in one of the wells.  Landau levels of a double-layer are,
therefore, degenerated (each level can be denoted by a well
index). The degeneracy in the well index is removed for lower
$B_{\parallel}$ (two layers gradually merge into a single-layer) and
for $B_{\parallel}=0$ each originally degenerated level is transformed
into an even/odd pair of Landau levels of the bonding subband.

The cyclotron mass corresponding to transitions between such a pair of
levels exhibit a singularity at high $B_{\parallel}$ (the transition
becomes dark there).

The energy separation of neighbouring levels from different pairs
first decreases with increasing $B_{\parallel}$, reaches minimum
around a critical field, and then returns to an original value.
Consequently, the corresponding cyclotron mass has a maximum at the
critical field. The transitions between such Landau levels are bright
in the high field limit, they corresponds to transitions between
Landau levels in individual quantum wells.
\begin{figure}[h]
\begin{center}\leavevmode
\includegraphics[angle=-90,width=0.95\linewidth]{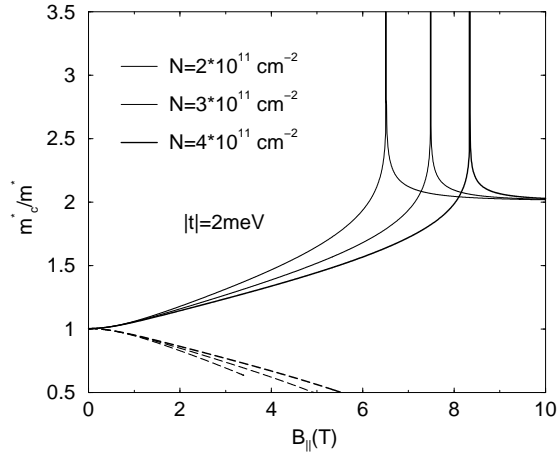}
\caption{The $B_{\parallel}$-dependence of the cyclotron effective
mass $m_c^*$ calculated quasi-classically.  The cyclotron effective
mass is normalized by the cyclotron effective mass at zero magnetic
field limit, $m^*$.  Full and dashed lines correspond to ground and
excited subbands, respectively.}
\label{fig06} 
\end{center} 
\end{figure}

The single-layer/double-layer transformation of electrons in the
bonding subband is even more clearly demonstrated in the
quasi-classical case.  For very small $B_{\perp}$ the number of
occupied Landau levels increases and it is more convenient to describe
the $B_{\parallel}$-dependence of the cyclotron mass in terms of the
Fermi energy, which separates the empty and occupied states, rather
than with a help of quantum numbers of empty and occupied Landau
levels.

The results are presented in Fig. \ref{fig06}. The cyclotron mass of
the bonding electrons exhibits a logarithmic singularity at the
critical field, above which the electrons cannot tunnel through
the barrier and form two insulated parallel electron sheets. The
growth of the electron concentration shifts the singularity position
to the side of larger in-plane fields.
\section{Conclusions}
We have performed a theoretical study of the cyclotron
resonance in 2D single-layers and double-layers subject to magnetic
fields of general orientation.

For strong $B_{\parallel}$ and $B_{\perp}$, the quantum-mechanical
approach applies: we have calculated the
field\discretionary{-}{-}{-}dependence of the energy levels and
transitions matrix elements between them, and found that $m^*_c$
becomes a function of both field components.

In strictly perpendicular magnetic fields only the intrasubband
transitions ($\delta m=0$, $\delta n=\pm 1$) are allowed. In 
tilted magnetic fields also the intersubband transitions ($\delta
m=\pm 1$, $\delta n=0$) become possible.
In general, the intrasubband transitions become dark and the
intersubband transitions bright at high $B_{\parallel}$.

The field dependence of the cyclotron mass in double well structures
is more complex than in the single wells.
\section{Acknowledgements}
This work was supported by AVOZ1-010-914 and  by the Grand
Agency of the Czech Republic under Grant No. 202/01/0754.                     
\enlargethispage*{0.5cm}

\end{document}